\documentclass[%
 reprint,
superscriptaddress,
 altaffilletter,
 amsmath,amssymb,
 aps,
pra,
]{revtex4-2}
\usepackage{graphicx}% Include figure files
\usepackage{dcolumn}% Align table columns on decimal point
\usepackage{bm}% bold math
\usepackage{textgreek}
\usepackage[mathlines]{lineno}% Enable numbering of text and display math
\usepackage{subfigure}
\begin{document}

\preprint{APS/123-QED}

\title{Imaging of single barium atoms in a second matrix site in solid xenon for barium tagging in a $^{136}$Xe double beta decay experiment}

%\date{\today}

%\roman{footnote}
\newcommand{\UCSD}{\affiliation{Physics Department, University of California San Diego, La Jolla, CA 92093, USA}}
\newcommand{\McGill}{\affiliation{Physics Department, McGill University, Montr\'eal, QC H3A 2T8, Canada}}
\newcommand{\Stanford}{\affiliation{Physics Department, Stanford University, Stanford, CA 94305, USA}}
\newcommand{\Erlangen}{\affiliation{Erlangen Centre for Astroparticle Physics (ECAP), Friedrich-Alexander University Erlangen-N{\"u}rnberg, Erlangen 91058, Germany}}
\newcommand{\PNNL}{\affiliation{Pacific Northwest National Laboratory, Richland, WA 99352, USA}}
\newcommand{\Carleton}{\affiliation{Department of Physics, Carleton University, Ottawa, ON K1S 5B6, Canada}}
\newcommand{\UMass}{\affiliation{Amherst Center for Fundamental Interactions and Physics Department, University of Massachusetts, Amherst, MA 01003, USA}}
\newcommand{\ITEP}{\affiliation{National Research Center ``Kurchatov Institute'', Moscow, 123182, Russia}}
\newcommand{\LLNL}{\affiliation{Lawrence Livermore National Laboratory, Livermore, CA 94550, USA}}
\newcommand{\UK}{\affiliation{Department of Physics and Astronomy, University of Kentucky, Lexington, KY 40506, USA}}
\newcommand{\BNL}{\affiliation{Brookhaven National Laboratory, Upton, NY 11973, USA}}
\newcommand{\SLAC}{\affiliation{SLAC National Accelerator Laboratory, Menlo Park, CA 94025, USA}}
\newcommand{\RPI}{\affiliation{Department of Physics, Applied Physics, and Astronomy, Rensselaer Polytechnic Institute, Troy, NY 12180, USA}}
\newcommand{\Laurentian}{\affiliation{School of Natural Sciences, Laurentian University, Sudbury, ON P3E 2C6, Canada}}
\newcommand{\IHEP}{\affiliation{Institute of High Energy Physics, Chinese Academy of Sciences, Beijing, 100049, China}}
\newcommand{\IME}{\affiliation{Institute of Microelectronics, Chinese Academy of Sciences, Beijing, 100029, China}}
\newcommand{\Sherbrooke}{\affiliation{Universit\'e de Sherbrooke, Sherbrooke, QC J1K 2R1, Canada}}
\newcommand{\Alabama}{\affiliation{Department of Physics and Astronomy, University of Alabama, Tuscaloosa, AL 35405, USA}}
\newcommand{\UNCW}{\affiliation{Department of Physics and Physical Oceanography, University of North Carolina Wilmington, Wilmington, NC 28403, USA}}
\newcommand{\UBC}{\affiliation{Department of Physics and Astronomy, University of British Columbia, Vancouver, BC V6T 1Z1, Canada}}
\newcommand{\TRIUMF}{\affiliation{TRIUMF, Vancouver, BC V6T 2A3, Canada}}
\newcommand{\Drexel}{\affiliation{Department of Physics, Drexel University, Philadelphia, PA 19104, USA}}
\newcommand{\CSU}{\affiliation{Physics Department, Colorado State University, Fort Collins, CO 80523, USA}}
\newcommand{\Yale}{\affiliation{Wright Laboratory, Department of Physics, Yale University, New Haven, CT 06511, USA}}
\newcommand{\USD}{\affiliation{Department of Physics, University of South Dakota, Vermillion, SD 57069, USA}}
\newcommand{\Mines}{\affiliation{Department of Physics, Colorado School of Mines, Golden, CO 80401, USA}}
\newcommand{\CUP}{\affiliation{IBS Center for Underground Physics, Daejeon, 34126, Korea}}
\newcommand{\UWC}{\affiliation{Department of Physics and Astronomy, University of the Western Cape, P/B X17 Bellville 7535, South Africa}}
\newcommand{\SUBATECH}{\affiliation{SUBATECH, Nantes Universit\'e, IMT Atlantique, CNRS/IN2P3, Nantes 44307, France}}
\newcommand{\Caltech}{\affiliation{California Institute of Technology, Pasadena, CA 91125, USA}}
\newcommand{\Bern}{\affiliation{LHEP, Albert Einstein Center, University of Bern, 3012 Bern, Switzerland}}
\newcommand{\Skyline}{\affiliation{Skyline College, San Bruno, CA 94066, USA}}
\newcommand{\SNOLAB}{\affiliation{SNOLAB, Lively, ON P3Y 1N2, Canada}}
\newcommand{\Muenster}{\affiliation{Institut f{\"u}r Kernphysik, Westf{\"a}lische Wilhelms-Universit{\"a}t M{\"u}nster, M{\"u}nster 48149, Germany}}
\newcommand{\FRIB}{\affiliation{Facility for Rare Isotope Beams, Michigan State University, East Lansing, MI 48824, USA}}
\newcommand{\Queens}{\affiliation{Department of Physics, Queen's University, Kingston, ON K7L 3N6, Canada}}
\newcommand{\Windsor}{\affiliation{Department of Physics, University of Windsor, Windsor, ON N9B 3P4, Canada}}
\newcommand{\TUM}{\affiliation{Physikdepartment and Excellence Cluster Universe, Technische Universit{\"a}t M{\"u}nchen, Garching 80805, Germany}}
\newcommand{\UCI}{\affiliation{Department of Physics and Astronomy, University of California, Irvine, Irvine, CA 92697, USA}}

\author{M.~Yvaine}\CSU
\author{D.~Fairbank}\CSU
\author{J.~Soderstrom}\CSU
\author{C.~Taylor}\CSU
\author{J.~Stanley}\CSU
\author{T.~Walton}\altaffiliation{Now at: Prism Computational Sciences, Madison, WI 53711, USA}\CSU
\author{C.~Chambers}\altaffiliation{Now at: TRIUMF, Vancouver, BC V6T 2A3, Canada}\McGill
\author{A.~Iverson}\CSU
\author{W.~Fairbank}\email{fairbank@colostate.edu}\CSU

\author{S.~Al Kharusi}\altaffiliation{Now at: Physics Department, Stanford University, Stanford, CA 94305, USA}\McGill
% ...

\date{\today}% It is always \today, today,

\begin{abstract}

Neutrinoless double beta decay is one of the most sensitive probes for new physics beyond the Standard Model of particle physics. One of the isotopes under investigation is $^{136}$Xe, which would double beta decay into $^{136}$Ba. Detecting the single $^{136}$Ba daughter provides a sort of ultimate tool in the discrimination against backgrounds. Previous work demonstrated the ability to perform single atom imaging of Ba atoms in a single-vacancy site of a solid xenon matrix. In this paper, the effort to identify signal from individual barium atoms is extended to Ba atoms in a hexa-vacancy site in the matrix and is achieved despite increased photobleaching in this site.  Abrupt fluorescence turn-off of a single Ba atom is also observed.  Significant recovery of fluorescence signal lost through photobleaching is demonstrated upon annealing of Ba deposits in the Xe ice. Following annealing, it is observed that Ba atoms in the hexa-vacancy site exhibit antibleaching while Ba atoms in the tetra-vacancy site exhibit bleaching. This may be evidence for a matrix site transfer upon laser excitation.  Our findings offer a path of continued research toward tagging of Ba daughters in all significant sites in solid xenon.

\end{abstract}

\maketitle

\section{Introduction}
Although neutrinos were discovered several decades ago, some of their most fundamental properties remain unknown. Neutrinoless double beta decay (0\textnu\textbeta\textbeta) is a sensitive probe for neutrino properties and new physics beyond the Standard Model of particle physics. If this hypothesized decay were observed, it would demonstrate a process that violates lepton number conservation and confirm that neutrinos are Majorana particles (that is, their own antiparticle). It would also  determine the absolute neutrino mass-scale, albeit with uncertainties of a theoretical nature \cite{Dolinski}. The EXO-200 experiment, utilizing a liquid xenon (LXe) time-projection chamber (TPC), made the first observation of two-neutrino double beta decay (2\textnu\textbeta\textbeta) in $^{136}$Xe and precisely measured its half-life at 2.165 ± 0.016 (stat) ± 0.059 (sys) x 10$^{21}$ yr \cite{EXO2011,EXO2014}. It also set a limit on the 0\textnu\textbeta\textbeta\, process in $^{136}$Xe of $T_{1/2}^{0\nu}$ $>$ 3.5 x 10$^{25}$ yr at a 90\% confidence level (CL) \cite{EXO2019}. A next-generation LXe experiment, nEXO, with 5 tonnes of enriched Xe, is expected to reach a half-life sensitivity to 0\textnu\textbeta\textbeta\, decay of 1.35 x 10$^{28}$ yr in a 10 year run \cite{nEXOSensitivity}. 

When a double beta decay event occurs in a $^{136}$Xe TPC, a $^{136}$Ba daughter, that could potentially be identified, is formed at the decay site. The detection of the daughter, referred to as barium tagging, could improve 0\textnu\textbeta\textbeta\, decay sensitivity by eliminating all backgrounds except those from 2\textnu\textbeta\textbeta\, \cite{Moe}. Implementation of barium tagging as a potential future upgrade to nEXO is being investigated by the collaboration. The elimination of all non-\textbeta\textbeta\ decay backgrounds in nEXO could increase the sensitivity by about a factor of 2-3 \cite{nEXOSensitivity}.

One Ba tagging effort in nEXO focuses on extraction from LXe and detection in solid xenon (SXe) \cite{Mong15, ChambersNature}, and another on extraction to gaseous Xe and detection in vacuum \cite{Brunner2015, Brunner}.  Capture of Ba$^{++}$ in a fluorescent molecule is also being investigated by the NEXT collaboration for barium tagging for a next-generation gaseous $^{136}$Xe double beta decay experiment  \cite{Nygren2016, McDonald2018, Byrnes2024}.

The principles of the method of Ba extraction in SXe and subsequent tagging have been described previously \cite{Mong15, ChambersNature}.  Briefly, a candidate 0\textnu\textbeta\textbeta \, decay event is identified by ionization signal characteristics such as total electron energy and multiplicity, and its location in the TPC is determined.  A cryoprobe is inserted into the TPC, the tip with a sapphire window is cooled below the Xe freezing point, and the potential $^{136}$Ba daughter is captured from the decay site in a small volume of solid xenon (SXe) on the window. The probe is extracted from the TPC volume and the LXe to a separate region where the SXe deposit can be cooled to a lower temperature and scanned for the presence of one or zero Ba atoms. Coincident observation of a Ba atom extracted from a candidate 0\textnu\textbeta\textbeta \, event site provides a positive confirmation of ββ decay.  

In LXe, the Ba\textsuperscript{++} daughter from the ββ decay is expected to quickly recombine to form Ba\textsuperscript{+} and mostly remain as such, as the band gap for LXe is less than the ionization potential for Ba\textsuperscript{+} \cite{Moe}. The best present analogue for estimating the final ion fraction for the Ba daughters in LXe is the 76(6)\% measured for \textsuperscript{214}Bi\textsuperscript{+} daughters of \textsuperscript{214}Pb beta decay \cite{EXO2015alpha}.  When Ba\textsuperscript{+} is captured into SXe from LXe at thermal energy, it is unclear what fraction will remain as Ba\textsuperscript{+} and what fraction will neutralize to Ba by capturing a stray electron. What is known is that both Ba atoms and Ba\textsuperscript{+} ions result from the deposition of 2000 eV Ba\textsuperscript{+} ions into SXe in vacuum \cite{Mong15}.  

In previous work, the spectroscopy of Ba atoms and Ba\textsuperscript{+} ions in SXe \cite{Mong15, McCaffrey2016, DavisMcCaffrey, Davis2018} and first results on imaging of single Ba atoms in SXe \cite{ChambersNature} have been presented. Theoretical work has identified the geometries of the matrix sites in SXe that give rise to some portions of the observed Ba absorption and emission spectra \cite{Davis2018, Buchachenko}. The 591 nm “blue site" emission (so named in Ref. \cite{Davis2018} by the order of absorption band wavelength) is associated with Ba atoms in tetra-vacancy (TV) site. The 577 nm emission peak has overlapping contributions from Ba atoms in two distinct sites. The 578 nm “violet site" emission is assigned to Ba atoms in asymmetric 5- or 7-atom vacancy (5/7V) sites \cite{Davis2018, Buchachenko}. The 577 nm “green site" emission, with an excitation peak between those of the violet site, is associated with Ba atoms in HV sites \cite{Davis2018, DavisMcCaffrey, Buchachenko}. The SXe matrix site of the weak 570 nm Ba emission peak is not yet identified. The site for the 619 nm emission is not addressed in published theoretical work, but is attributed to a single-vacancy (SV) site \cite{ChambersNature, Gervais}. One possible formation mechanism for this less energetically favorable site in our work is the capture of a Ba\textsuperscript{+} ion in a SV site and its later neutralization by recombination with a free electron.  The predominant emission lines of Ba\textsuperscript{+} in SXe have also been found experimentally \cite{Mong15}, but to date no theoretical results on Ba\textsuperscript{+} in SXe have been published.

First images of single Ba atoms in SXe  were achieved with SV site Ba atoms emitting at 619 nm \cite{ChambersNature}.  For high efficiency Ba tagging, it is important to extend single Ba imaging from the SV site to all Ba matrix sites in SXe and to also demonstrate single Ba\textsuperscript{+} imaging in SXe.  In this paper, the first step in this extension is presented, imaging of single Ba atoms in a second site, the HV matrix site in SXe using the 577 nm emission peak. Because a Ba atom in this matrix site photobleaches more rapidly than in the SV site \cite{Mong15}, single atom imaging is more challenging.

In parallel work, imaging of single rubidium atoms in a solid neon matrix by laser induced fluorescence was reported recently \cite{Weinstein2024}. A unique aspect of that work is the ability to optically pump and read out the spin state of the atom.    Single atom imaging of \textsuperscript{25}Mg in solid neon has also been proposed for measurement of an extremely low nuclear reaction cross section of astrophysical importance \cite{Singh2019}.

\section{Experimental Setup}

\begin{figure}
    \includegraphics[width=8cm]{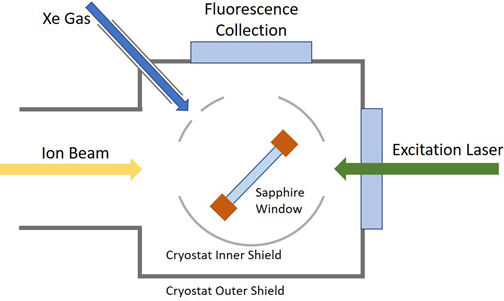}
\caption{A schematic diagram of the sample apparatus.  The excitation laser is opposite the Ba\textsuperscript{+} ion beam, and the fluorescence is collected in a perpendicular direction.  Xenon gas is directed toward the window by a tube that enters the vacuum behind the plane of the diagram.}
\label{fig:apparatus}
\end{figure}

The apparatus used in this work is similar to that used for imaging single Ba atoms in SV sites of SXe \cite{ChambersNature}. A schematic diagram of the sample region is shown in Fig. \ref{fig:apparatus}. Matrix samples for the imaging experiments are created on a sapphire window of 0.75 inch diameter \cite{Meller} that is mounted on the cold finger of a cryostat at a temperature of 50 K. Near the window an inlet tube provides gaseous xenon flow that freezes on the window to create the matrix sample.  In Ba\textsuperscript{+} deposits, pulses of Ba\textsuperscript{+} ions are directed towards the center of the window during the deposition of xenon. The ion pulses begin about 5 seconds after the start of xenon flow and are followed by \url{~}5 s of xenon deposit. During implantation, some of the Ba\textsuperscript{+} ions recombine with the \url{~} 20 free electrons that result from ionization of the SXe matrix by each 2000 eV Ba\textsuperscript{+} ion.  Thus, a sparse layer of Ba atoms and Ba\textsuperscript{+} ions remains in the middle of a SXe layer of thickness \url{~}0.5 µm.  After deposition, the window and sample are cooled to 12 K for imaging experiments.  The window plane is tilted at an angle of about 45° relative to the ion beam so that sample creation, excitation, and fluorescence collection can be done without adjusting the position of the window and the optics.

For the imaging of single Ba atoms, a Coherent 599 dye laser is used to illuminate the sample from a direction opposite the ion beam. A  laser wavelength of 565.0(1) nm is chosen for near-optimal excitation of HV site Ba atoms and negligible excitation of those in the 5/7V site \cite{DavisThesis, Davis2018}.  The typical laser power is 0.4 µW, about two orders of magnitude less than that used for imaging Ba atoms in the SV site, to accommodate the faster fluorescence bleaching of HV site Ba atoms. The laser is focused at the window front surface (toward the ion beam) using an aspheric lens of 7.9 cm focal length to a radius, w, of $\sim 3$ µm. An angled optical flat is used for compensation of the astigmatism of the tilted window.

The resulting fluorescence is collected and collimated by a f/1.4 50 mm focal length NIKKOR camera lens mounted directly above the window. The light is filtered by a Semrock sharp-edged 582 nm bandpass filter with 21 nm FWHM bandpass and an additional tilted 568nm long pass filter for greater blocking of stray laser light. The combined bandpass transmits the 577 nm peak with high efficiency and some of the 590 nm peak.  Because the latter peak has greater bleaching at excitation wavelength 565 nm \cite{Mong15},  the contribution of TV Ba atoms to the fluorescence signal is small.  The collected fluorescence is steered by two alignment mirrors into a 200 mm focal length camera lens, which focuses the light into an Andor iXON 3 CCD camera. This model features EM (electron multiplying) gain for single photon counting and is held at -100 °C. For the experiments reported in this paper, CCD settings of EM gain 300 and preamp gain 1 were used. The conversion factor from counts to detected photons is 12.8 counts/photoelectron, determined by experimental scaling from factory calibration with no EM gain. This value is used in this work. 
 An alternate confirmation of this calibration from the measured slope of the exponential distribution of single photoelectron pulse heights is \url{~}15 counts/photoelectron.

Movements of the sample window with an amplitude of \url{~}13 µm occur during pulses of pressurized helium in the cryostat, so the laser is shuttered with \url{~}40\% duty cycle to minimize this effect. The residual blurring of the laser position relative to single Ba atoms was determined from analysis of images of single Ba atoms in SV sites to be equivalent to effective laser radii of w\textsubscript{x}= 5.73 ± 0.26 µm and w\textsubscript{y}= 6.62 ± 0.40 µm \cite{ChambersThesis}.

The ion source is a commercial Colutron ion gun with E$\times$B mass filter for selecting singly ionized Barium ions at 2000 eV \cite{Colutron}.  A series of deflection plates and Einzel lenses provides steering and focusing to deposit the Ba\textsuperscript{+} ions in the center of the window. Typical currents measured by an extractable Faraday cup close to the window were on the order of 5-10 nA. To deposit a small number of ions in a matrix sample, a pair of pulsing plates are set at ±200 V to deflect the beam away from the window. These plates are set to 0 V during pulse periods of \url{~}1.5 µs to create ion bunches. The induction signal in a set of three disks through which the ions pass provides a monitor of the pulse size during deposits. The effective area of the ion beam was determined to be 6.7 mm$^{2}$ prior to the experiments by replacing the window with a Faraday cup.  The uncertainty on ion beam density at the window is +20\% and -30\% \cite{ChambersThesis}.

The vacuum was provided by a small turbo pump on a CF2.75" port attached near the cryostat and two larger turbo pumps on CF8" ports attached to the ion beam system. Residual gas levels monitored with a SRS residual gas analyzer are typically around 1x10$^{-7}$ Torr before the cryostat is turned on. Flow of research grade xenon gas (99.995\% purity) is controlled with a needle valve for a desired SXe growth rate that has been calibrated by interference fringes. No further purification is used due to the small flow rates used for sample creation. 

As in our previous work, to image single Ba atoms we raster scan the focused  laser over the sample in steps of $4~\mathrm{\mu m}$   (transverse to the laser and ion beams) and identify spatially resolved Ba peaks through their fluorescence response. For each laser scan step, the fluorescence signal in counts is determined by summing the counts in a 5$\times$5 pixel region centered at the laser position and subtracting a digitizing offset (pedestal) in the CCD readout electronics of 100 counts/pixel.  A plot of net counts vs. laser x,y position is called a composite image and shows where single Ba atoms are located.

In some of our experiments, the output coupler of the dye laser was adjusted between deposits.  This resulted in a shift in the starting y position of some scans of up to 6 laser steps relative to previous deposits.  There were no x position shifts.

\section{Results}

\begin{figure}
    \includegraphics[width=0.4\textwidth]{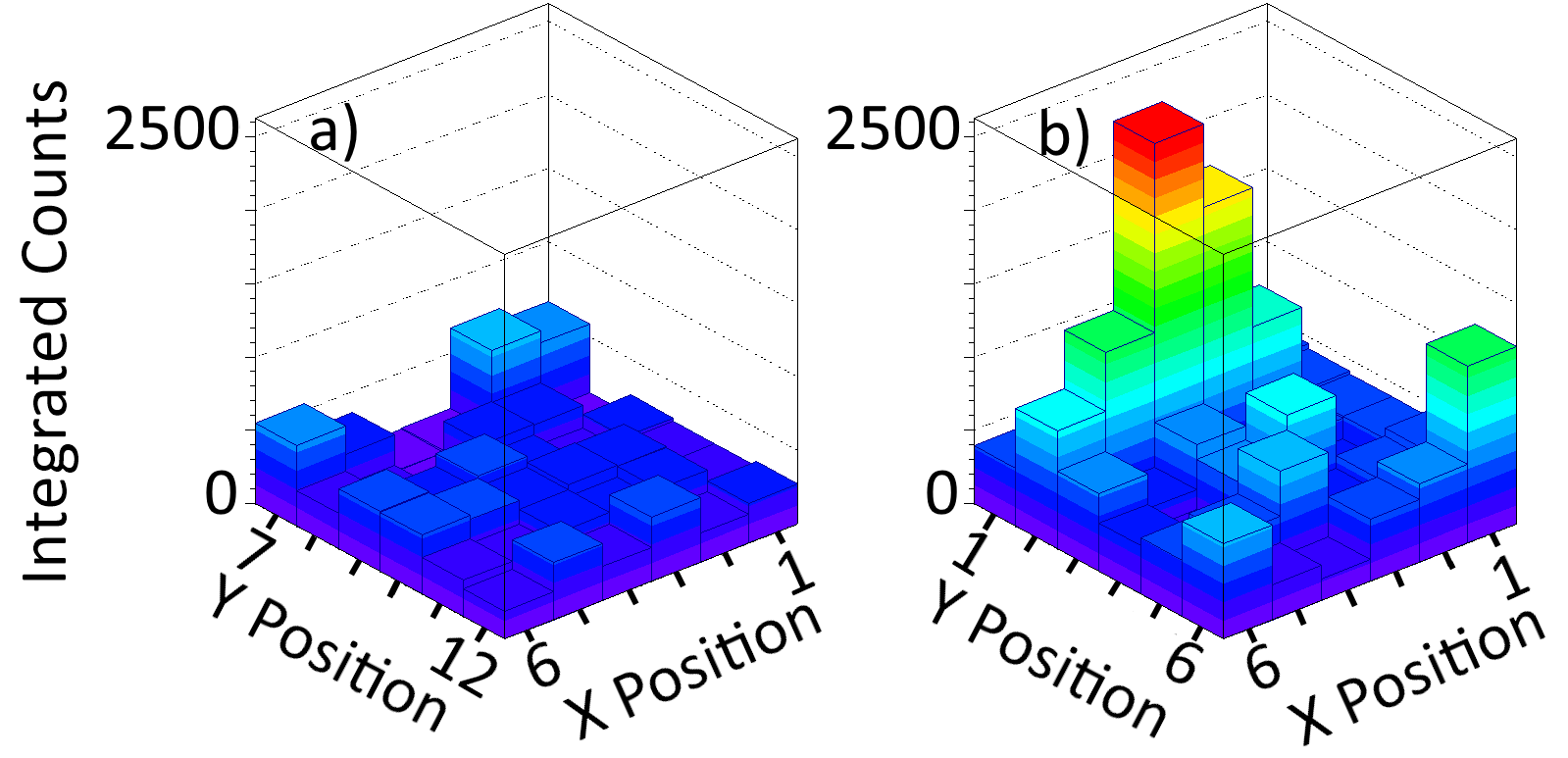}
    \includegraphics[width=0.4\textwidth]{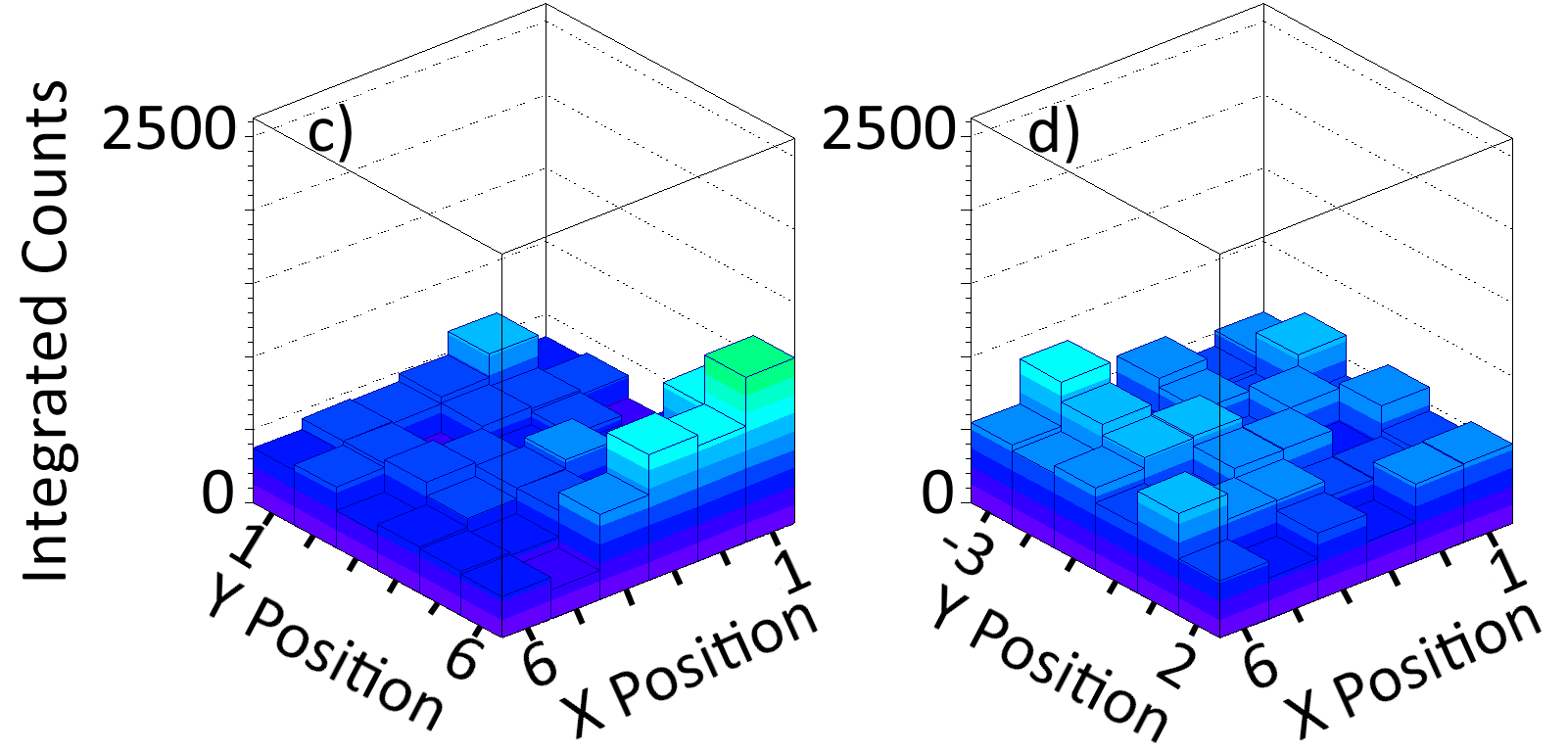}
\caption{A set of lego plots of integrated fluorescence counts vs. xy laser position in successive scans from a) to d): a) SXe-only deposit; b) new SXe with barium deposit, with one distinct peak at laser position (4,2) of the scan; c) rescan of the SXe with barium deposit in which the prominent feature has photobleached; d) new SXe-only after the previous deposit was evaporated.  The axes in these plots are rotated 90° clockwise compared to plots shown later in this manuscript for better visibility.}
\label{fig:scan1}
\end{figure}

\begin{figure*}
    \includegraphics[width=1\textwidth]{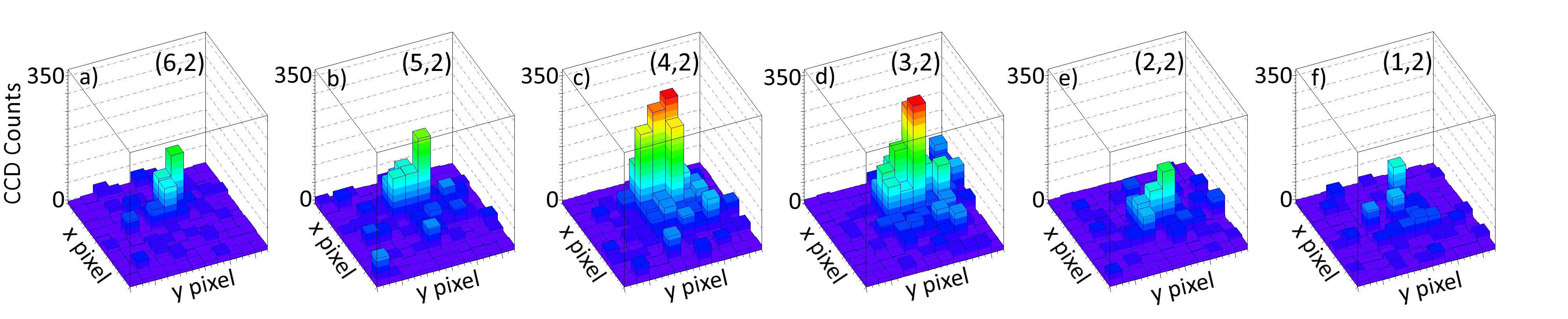}
\caption{A set of lego plots of successive $15\times15$ pixel ccd images from the scan depicted in Fig. \ref{fig:scan1} (b). The laser position in scan steps is given in parentheses at the top right of each image.  The CCD counts are pedestal subtracted.  In the first four images, as the laser passes over the single atom, the fluorescent signal increases, but remains in the same position as the laser moves across. This is an indicator that the fluorescence is coming from a single location in the scan region, rather than being a feature of the moving laser.}
\label{fig:scan1frames}
\end{figure*}

Results for one set of $6\times6$ step scans with 0.4 µW of laser power for 10 s (25 $s\times40$\% duty cycle) and peak exposure 67 nWs/µm$^{2}$ per position are shown in Fig. \ref{fig:scan1}.  Summed fluorescence counts vs. laser x,y position are shown for each scan.  In (a), a scan over a fresh solid xenon deposit exhibits little signal.  In the first two laser positions (1,7) and (2,7), extra background light from the filament of the residual gas analyzer was present, which was turned off by the third position.  The average background for the other 34 positions is 281 counts with a standard deviation of 102 counts.  

After the Xe-only deposit was evaporated at 100 K, a new Xe deposit with 57$_{-17}^{+12}$ Ba\textsuperscript{+} ions per scan area (576 µm$^{2}$) was deposited at 50 K.  In the composite image (b) a single Ba peak stands out clearly above the xenon background with a peak of \url{~}2600 counts in the frame with the largest signal. The raw images of the laser positions in the second row, (6,2) through (1,2), are displayed in Fig. \ref{fig:scan1frames}. As the laser moves over the Ba atom, the signal dramatically increases, until vanishing as the laser moves away from that position.  

In a second scan of this sample, Fig. \ref{fig:scan1} (c), fluorescence of the single Ba atom is absent, indicating a bleaching of this atom’s fluorescence during the exposure of the first scan. It is not clear whether or not the higher signal around laser position (1,6) in scan (b) and rescan (c) of Fig. \ref{fig:scan1} is from weak excitation of another Ba atom that was just outside the scan area. After the second scan, the deposit was evaporated and a new xenon-only deposit was made and scanned.  No significant fluorescence peaks were observed in this SXe-only scan, Fig. \ref{fig:scan1} (d).

\begin{figure}
    \includegraphics[width=0.45\textwidth]{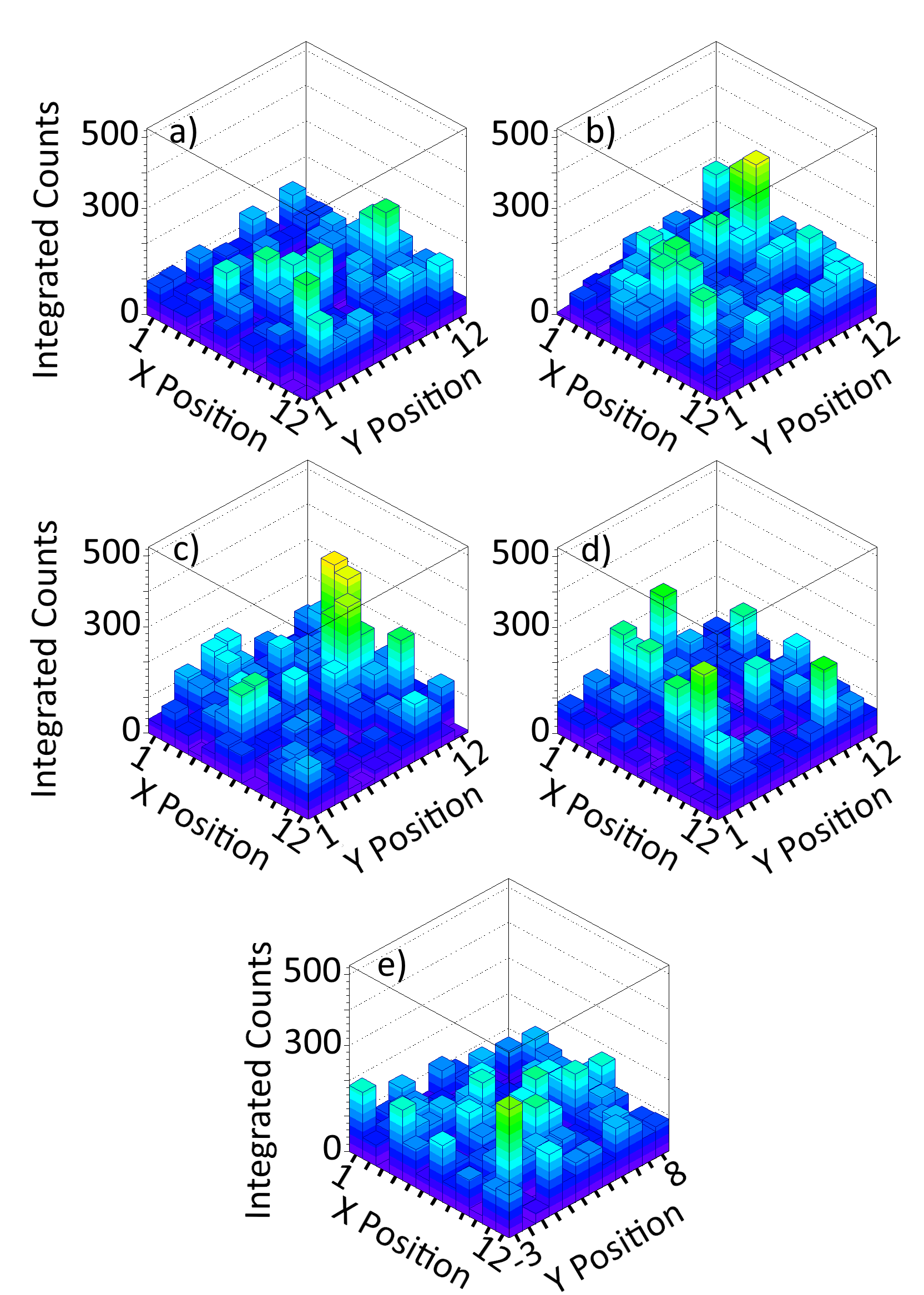}
\caption{Lego plots of a successive set of deposits and composite scans with lower exposure time: a) SXe-only deposit and scan; b) SXe deposit with barium which has a few higher features compared to (a); c) rescan of the same deposit in which the strongest feature in (b) persists; d) another rescan of the same deposit as in (b) and (c) after exposure for 50 frames on the feature of interest at position (6,10); e) SXe-only deposit with a similar average signal level as in (a). A laser alignment between scans (d) and (e) resulted in a -4 step y-position shift.  }
\label{fig:scan2}
\end{figure}

Other experiments were done with the same laser power, 0.4 µW, but reduced laser exposure time,  2.5 s per position, to reduce the amount of photobleaching of the candidate peaks during a scan. One purpose of these experiments was the possible observation of the fluorescence turnoff of one Ba atom.

The composite $12\times12$ step scan images of a Xe-only deposit and a subsequent set of scans of a Ba deposit in SXe are shown in Fig. \ref{fig:scan2}. Lower exposure resulted in fewer photons detected, 6 on average per position in the Xe-only scan (a), resulting in higher relative statistical fluctuations. After the scan, the Xe-only deposit was evaporated and another deposit of xenon with 326$_{-98}^{+65}$ Ba\textsuperscript{+} ions in the scan area (2304 µm$^{2}$) was made, scanned and then rescanned.  Results are shown in (b) and (c). A potential single Ba atom candidate is present in both the Ba scan (b) and rescan (c) at position (6,10) with a peak of \url{~} 360 counts in (b) and \url{~} 400 counts in (c).

\begin{figure}
    \includegraphics[width=0.4\textwidth]{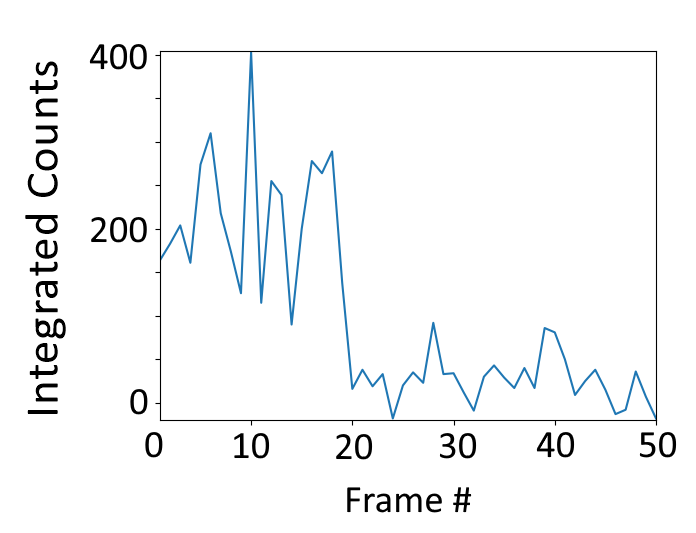}
\caption{Signal vs. time from 50 frames with the laser at position (6,10) of the Ba feature that persisted in Fig. \ref{fig:scan2} (b) and (c).  Each frame had 2.5 s of 0.4 µW laser power.}
\label{fig:spotsit}
\end{figure}

Since the fluorescence signal of this potential single Ba atom candidate persisted on rescan, the laser was then moved to position (6,10). 50 CCD frames were taken in this location at the same level of exposure as in the scan. The integrated signal for each of the 50 frames is shown in Fig. \ref{fig:spotsit}. For the first 19 exposures there was an average of 215±18 counts detected per frame; then the signal abruptly dropped to an average of 26±5 counts detected per frame. Such a turn-off to background level is a strong indicator of fluorescence from a single Ba atom. In a subsequent rescan of the area, Fig. \ref{fig:scan2} (d), the Ba atom signal was absent at position (6,10), as expected.  After this deposit was evaporated, another Xe-only deposit was made and imaged in Fig. \ref{fig:scan2} (e). The average signal had around 7 detected photons per position, only slightly higher than in the first Xe only deposit of Fig. \ref{fig:scan2} (a).  Both Xe-only scans are visibly lower in signal and variations than the three scans of the deposit with barium.  

\begin{figure}
    \centering
        \subfigure{\includegraphics[width=0.225\textwidth]{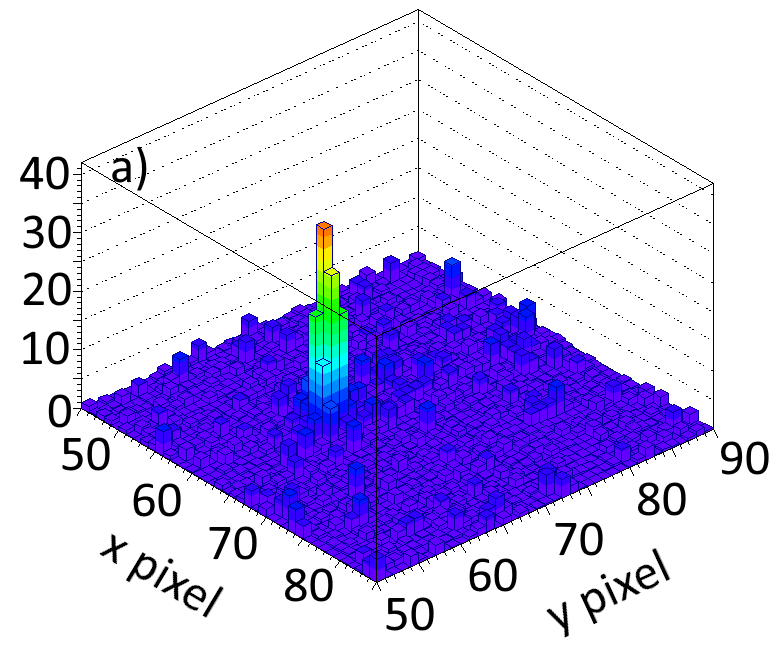}}
    \hfill
        \subfigure{\includegraphics[width=0.225\textwidth]{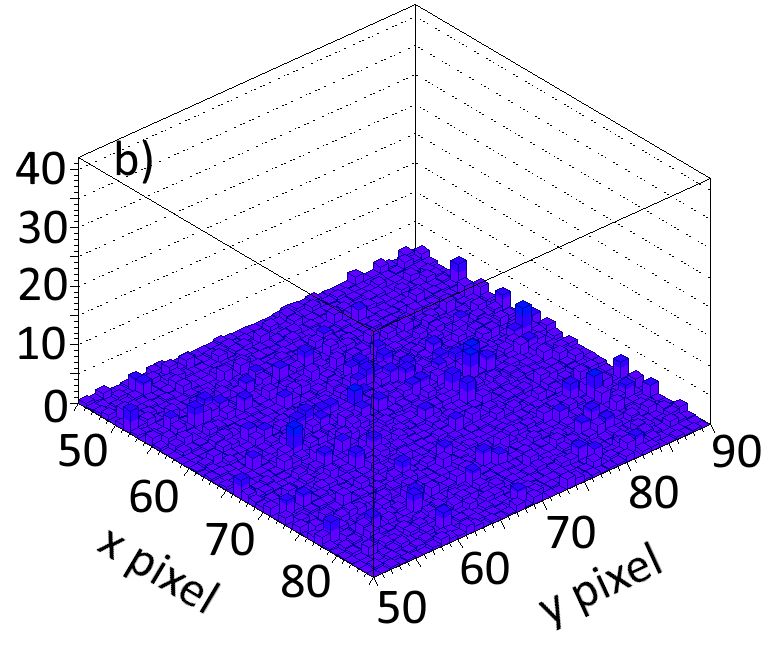}}
    \hfill
        \subfigure{\includegraphics[width=0.48\textwidth]{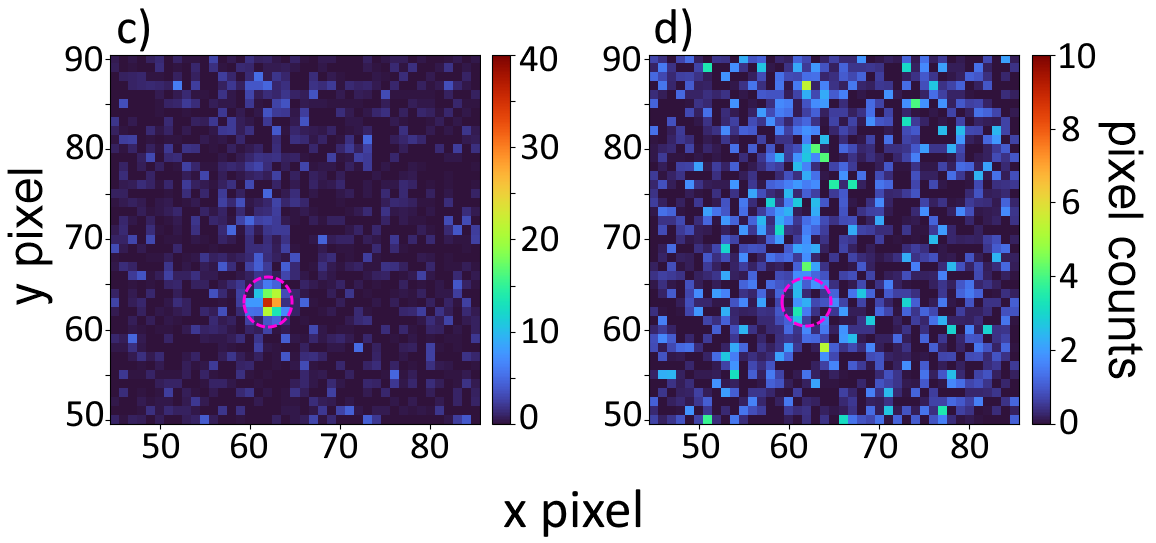}}
    \caption{Lego plots of a) the average of the first 19 frames of Fig. \ref{fig:spotsit} at position (6,10) (fluorescence "on") and b) the average of the remaining frames (fluorescence "off"); 2D intensity plots show laser position: c) first 19 frames and d) remaining frames rescaled 4x.}
    \label{fig:spotsitimage}
\end{figure}

 The apparent discrepancy of \url{~} 2 photons/frame “off" background level in Fig. \ref{fig:spotsit} with  6-7 photons/frame background in the Xe-only scans of Fig. \ref{fig:scan2} (a) and (e) is also found in other experiments.  For example, in four Xe-only scans and seven “spot sits" with fixed laser position on this day and the next day, the spot-sit “off" signals were consistently smaller and \url{~} 40\url{%} of the Xe-only averages.  A possible mechanism is partial photobleaching of the background that is not observed in individual frames due to low photon statistics.

Lego plots of averages of the raw pedestal-subtracted CCD images of first 19 frames (“on") of the run in Fig. \ref{fig:spotsit} and of the latter 31 frames (“off") are shown in Fig. \ref{fig:spotsitimage} (a) and Fig. \ref{fig:spotsitimage} (b), respectively.  A clear image of a single Ba atom is seen in (a). To confirm that the turnoff of atom fluorescence is not as a result of laser movement, two-dimensional plots of these averages are shown in Fig. \ref{fig:spotsitimage} (c) and (d), with a 4x magnified intensity scale in (d).  The small background signal from surface fluorescence in the latter 31 frames lies at the lower end of a weak line of fluorescence from a very low concentration of Cr$^{3+}$ ions in the bulk of the sapphire window \cite{WaltonThesis}.  The surface background is at the same position within 0.5 pixel as the laser fluorescence signal of the former 19 frames (marked by a circle).  This confirms the absence of xy-motion of the laser.  Technically, xy-motion of the window in its plane that carries the Ba atom out of the laser beam, cannot be excluded by this method.  However, the sudden turnoff and the absence of an atom signal at a shifted position in the rescan strongly suggest that the window did not move.

\section{Discussion}

A comparison of the images obtained of SV site Ba atoms \cite{ChambersNature} and those in HV sites in this work illustrates the need for characterizing, understanding and mitigating photobleaching.  Theoretical understanding is difficult in part because many absorption-emission cycles (10$^{3}$ to \textgreater10$^{7}$) occur before some bleaching mechanism terminates the process.  In addition, in multivacancy sites in SXe, the smaller excited $^{1}$P state Ba atom likely wanders around the larger SXe hole in the few ns before it emits, as was found in simulations of excited Na atoms in TV sites in solid krypton \cite{Ryan2010}.

Reversible fluorescence wavelength change due to transfer between thermally stable matrix configurations has been observed for Yb in SAr \cite{Tao2015}. Site transfer has not yet been identified as a major bleaching mechanism for Ba in the HV site, although first evidence for transfer from the TV site to the HV site is presented below.  Because recovery in dark from bleaching has not been observed, even over hundreds of seconds, optical pumping to metastable D or $^{3}$P\textsubscript{0,2} states, as would occur in vacuum, is not a bleaching mechanism.  In a non-symmetric diatomic BaXe environment, metastable state lifetimes are significantly reduced because the parity selection rule of atoms is modified by Stark mixing \cite{Buchachenko2}. Chemical reactions with rare impurities in the matrix are minimized by forming the matrix at 50 K, which is above the condensation temperature of  N\textsubscript{2}, O\textsubscript{2} and H\textsubscript{2} in vacuum. Recovery by modest annealing, as discussed below, rules out this as the dominant bleaching mechanism.

\begin{figure}
    
    \includegraphics[width=0.45\textwidth]{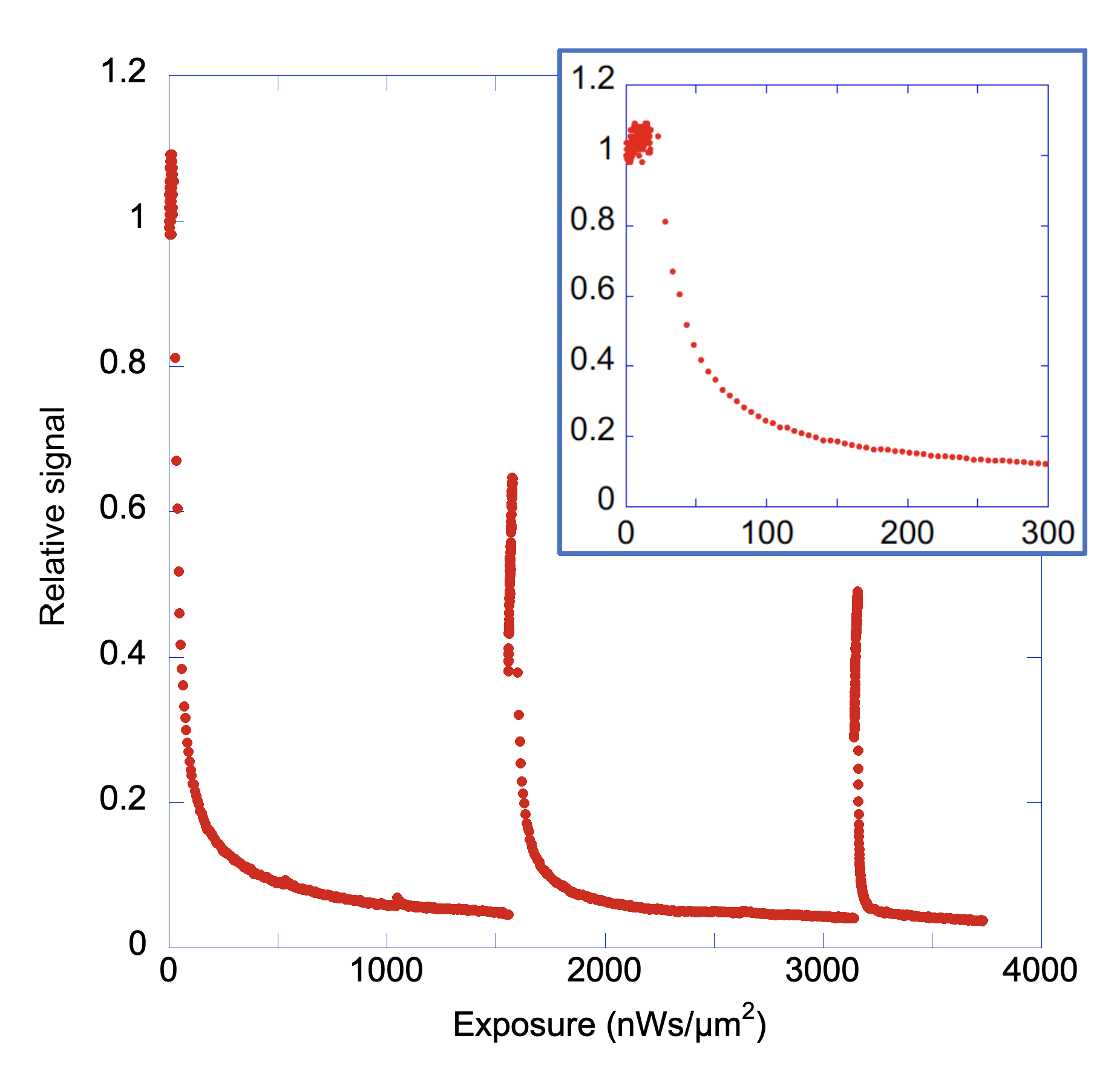}
    \caption{Normalized 577 nm Ba fluorescence vs. exposure for a large Ba deposit.  Recovery of fluorescence during two subsequent sample anneals to 40 K in the dark can be seen.  The inset shows an expanded view of initial bleaching of this Ba deposit.}
    \label{fig:bleaching}
\end{figure}

Results of a bleaching experiment with a large Ba deposit, \url{~} 5 s of DC Ba$^+$ beam, and a defocused laser beam at 565 nm with w\textsubscript{x}=258 µm and w\textsubscript{y}=315 µm from a continuous-wave optical parametric oscillator (OPO) are shown in Fig. \ref{fig:bleaching}.  Fluorescence was passed through a spectrometer to produce successive Ba emission spectra. The combination of a narrow spectrometer slit compared to w\textsubscript{x} and a narrow y integration range compared to w\textsubscript{y} assures uniform laser intensity in the region sampled.  As shown in the inset, in the first 100 frames,  with 0.09 nW/µm$^{2}$ laser intensity and 2 s exposure time (18 nWs/µm$^{2}$), a slight antibleaching occurs in the 577 nm fluorescence peak of HV site Ba atoms.  During this period, the 590 nm fluorescence peak for TV site Ba atoms bleaches by about a factor of 2.  The next 300 frames were taken with 1 s exposure time and a laser intensity 5.1 nW/µm$^{2}$ that is close to the central intensity of the focused laser beam in the experiments of Figs. \ref{fig:scan1} - \ref{fig:spotsitimage}, 6.7 nW/µm$^{2}$.  The fluorescence in these frames decays non-exponentially with laser exposure but can be described by a power law with exponent -0.6. At the additional exposure per frame of the scans in Figs. \ref{fig:scan1} - \ref{fig:scan1frames}, 67 nWs/µm$^{2}$, about 1/4 of the Ba fluorescence remains in Fig. \ref{fig:bleaching}.  If this curve can be interpreted as individual Ba atoms turning off as in Fig. \ref{fig:spotsit} with varying times, then, on average, 1/4 of the fluorescing HV Ba atoms in the center of the focused laser beam could contribute a full fluorescence signal for one scan frame at this exposure.  For a frame of the lower exposure scans in Fig.~\ref{fig:scan2} (17 nWs/µm$^{2}$) more than 1/2 of the fluorescing HV Ba atoms could fluoresce for a full frame's maximum exposure. Apparently, the Ba atom in Figs. \ref{fig:spotsit} and \ref{fig:spotsitimage} whose fluorescence persisted through scan, rescan and 19 frames with total exposure \url{~} 350 nWs/µm$^{2}$ was a particularly long fluorescing atom with about 10\url{%} probability.

\begin{figure}
   
    \includegraphics[width=0.45\textwidth]{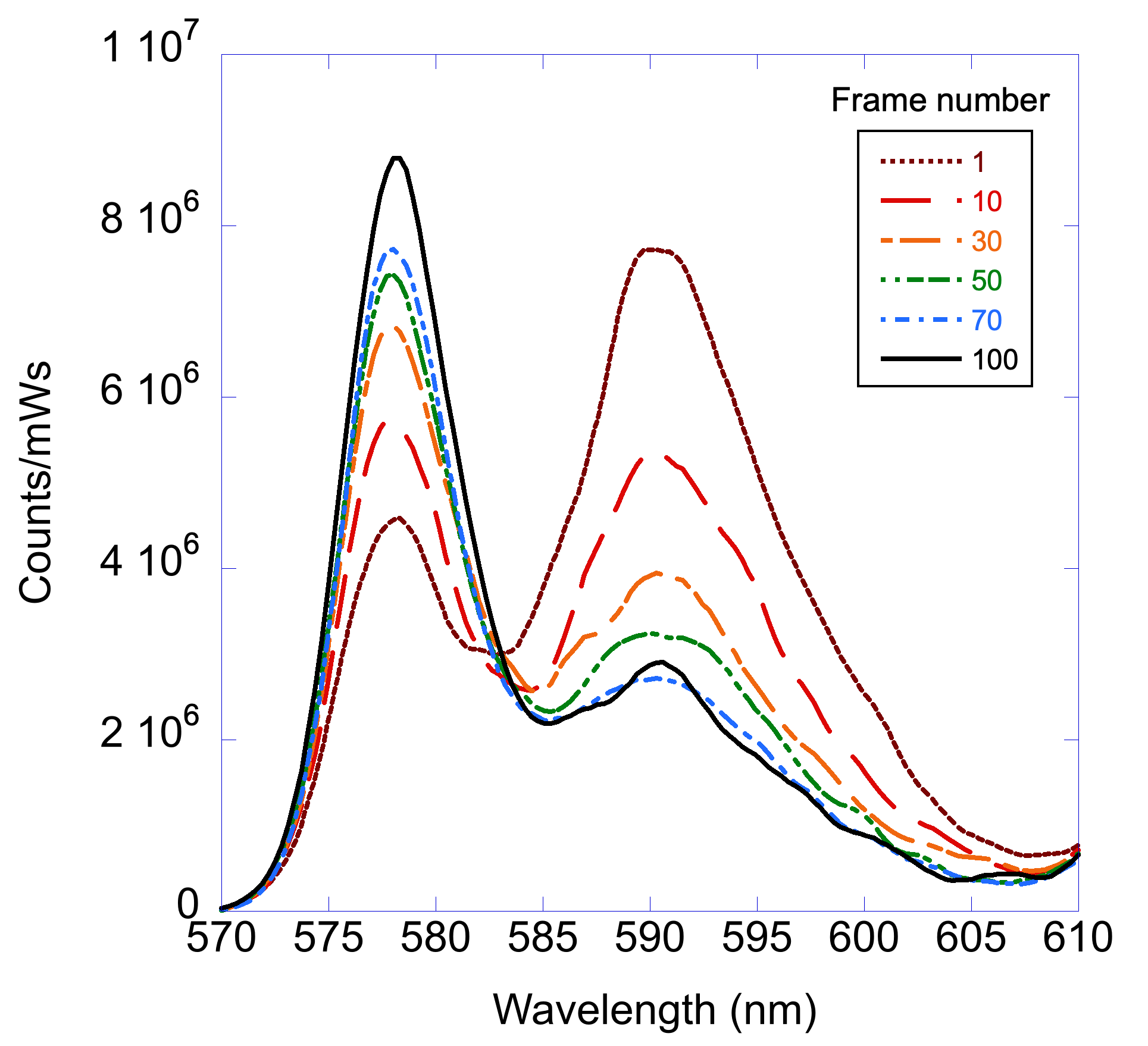}
    \caption{Selected spectra in the first 100 frames after the first anneal showing anti-bleaching of the 577 nm peak and concurrent bleaching of the 590 nm peak.}
    \label{fig:antibleaching}
    
\end{figure}

Recovery from excitation-induced transfer from one matrix site to another by annealing has been reported for Yb and K in SAr and Ag in SXe \cite{Tao2015,Kometer1996,Steinmetz1987}.  Such a process is also observed for Ba in SXe in Fig.~\ref{fig:bleaching}.  Following the bleaching run in the first 1559 nWs/µm$^{2}$ of Fig. \ref{fig:bleaching}, the sample was annealed in the dark to 40 K and returned to 12 K over 10 minutes. An immediate recovery to about 40\% of the original 577 nm signal was observed.  Over 100 low intensity frames, antibleaching of 577 nm increased the recovery to 65\% while the 590 nm fluorescence bleached significantly. Sample spectra from this run are shown in Fig.~\ref{fig:antibleaching}.  In the runs to this point the x- and y-positions of the OPO laser were steady as monitored by the y-position and intensity of the 619 nm peak for SV site Ba atoms that bleach little under these exposures.  Thus for these cases the recovery is not due to movement of the laser to less bleached Ba atoms.  After further bleaching, as much as half of the original signal returned on a second annealing cycle while the sample was not exposed.  During the second bleach, a little flattening of the bleaching curve is observed at the bottom due to y-movement of the laser beam at that time of magnitude 0.4 times the laser radius w. Otherwise, laser motion did not significantly change the exposure in these runs.  Further investigation is planned to see if annealing could be utilized to extend fluorescence duration and increase the number of detected photons per atom.  This could enhance the efficiency and validation for imaging of single Ba atoms in HV sites.

\section{Conclusions}

A key technology in a Ba tagging scheme for $^{136}$Xe neutrinoless double beta decay searches is the ability to identify and count single atoms in solid xenon. While single Ba imaging was previously demonstrated for barium atoms in SV sites in solid xenon, the ability to do this for Ba atoms in other matrix sites that experience greater photobleaching is critical. The first such images of single Ba atoms in the HV site are presented in this work. Sudden fluorescence turnoff of one Ba atom with particularly long fluorescence duration was also observed.

Thus far, the percentage of Ba$^+$ ions deposited into SXe in vacuum and imaged as single Ba atoms is in the few percent range for both SV and HV sites.  This ratio includes an unknown fraction of Ba$^+$ ions neutralized to Ba in present experimental conditions.  In the former case, a limiting factor may be low population of the energetically disfavored SV site.  For Ba atoms in the HV site a method to overcome photobleaching would be helpful.  The observed 2/3 recovery of fluorescence for the 577 nm line by annealing is a promising step in this direction.  The conversion of Ba atoms from the faster bleaching TV site of SXe to the HV site, in which single Ba imaging has been demonstrated here, is a hopeful avenue to explore further toward counting of the Ba atoms that are initially in TV sites.  The SV, TV, HV and 5/7V sites constitute the major configurations of Ba in SXe observed in our work.  Research is also underway in our laboratory on single Ba$^+$ imaging in solid xenon.  In liquid xenon, it is not known into which of these sites the daughter Ba atoms and Ba$^+$ ions from double beta decay will be captured when frozen into SXe. Initial studies are ongoing in our group.  Measurement of the Ba-ion fraction following radioactive beta decay of Cs, for example, \textsuperscript{136}Cs to \textsuperscript{136}Ba, \textsuperscript{138}Cs to \textsuperscript{138}Ba or \textsuperscript{139}Cs to \textsuperscript{139}Ba, would provide additional insights into the expected Ba/Ba$^+$ fraction following a double beta decay.  Ultimately, efficient single Ba and Ba$^+$ imaging in all sites is desirable to investigate.  In this regard, a significant step forward has been presented in this work.

\section{Acknowledgements}

The authors are grateful to Jaret Stickrod and Eli Bayarsaikhan for assistance with experiments and to John McCaffrey, Alexei Buchachenko and Benoit Gervais for helpful discussions of the theory of Ba atoms in noble gas matrix.  This material is based upon work supported by the National Science Foundation under Grants No. 2011948 and 1649324. The authors gratefully acknowledge support for nEXO from the Office of Nuclear Physics within DOE's Office of Science, and NSF in the United States; from NSERC, CFI, FRQNT, NRC, and the McDonald Institute (CFREF) in Canada; from IBS in Korea; and from CAS and NSFC in China. This work was supported in part by Laboratory Directed Research and Development (LDRD) programs at Brookhaven National Laboratory (BNL), Lawrence Livermore National Laboratory (LLNL), Oak Ridge National Laboratory (ORNL), Pacific Northwest National Laboratory (PNNL), and SLAC National Accelerator Laboratory.

\section{References}

\bibliographystyle{unsrt}
\bibliography{references}

\end{document}